# Simultaneous real-time multispectral fluorescence and reflectance imaging for enhanced intraoperative guidance

KYRIAKOS PENTARAKIS[1,2,*] GEORGE KAKAVELAKIS[1] CONSTANTINOS PETRIDIS[1] AND GEORGE THEMELIS[2]

[1]*Department of Electronic Engineering, School of Engineering, Hellenic Mediterranean University, Romanou 3, Chalepa, Chania, Crete GR-73100, Greece*
[2]*Leica Microsystems (Schweiz) AG, Max Schmidheiny-Strasse 201, Heerbrugg 9435, Switzerland*
*\*kyriakos.pentarakis@leica-microsystems.com*

**Abstract:** Intraoperative optical imaging is essential for surgical precision and patient safety, but current systems present anatomical and fluorescence information separately, causing delays and increasing cognitive load. A unified system for simultaneous visualization is critically needed to enhance guidance and efficiency.

To develop and validate a multispectral imaging system that overcomes fragmented intraoperative imaging by simultaneously capturing and integrating white light reflectance and multiple fluorescence signals in real-time, enhancing surgical precision and safety.

The system integrates two synchronized color cameras with optimized optical components, including custom filters and a beam splitter, tailored for clinically relevant fluorophores—protoporphyrin IX (PpIX), fluorescein sodium, and indocyanine green (ICG). Advanced image processing techniques enhance visualization accuracy. Performance was evaluated using phantoms simulating clinical conditions.

The system provided real-time, simultaneous visualization of white light and multiple fluorescence signals without mode switching. Fluorescence emissions from PpIX, fluorescein sodium, and ICG were accurately separated using linear unmixing algorithms. Reflectance images showed high color fidelity, with Delta E (ΔE) values indicating imperceptible to barely perceptible differences compared to a conventional camera. Latency was minimal, ensuring immediate visual feedback suitable for intraoperative use.

This multispectral imaging system addresses the limitations of fragmented intraoperative imaging by enabling continuous, real-time, integrated visualization of anatomical and fluorescence information. It streamlines workflow, reduces cognitive load, and supports more precise interventions. By leveraging components similar to conventional surgical microscopes, it offers a practical, cost-effective solution aligned with clinical needs. Future work includes clinical validation and advanced multi-fluorophore analysis to further improve outcomes.

**Keywords:** Multispectral imaging, fluorescence imaging, fluorescence-guidance surgery, tumor surgery, vascular surgery.

## 1. Introduction

Intraoperative optical imaging has transformed modern surgical practice by enabling real-time, high-resolution assessment of both anatomical structures and pathological conditions [1]. White-light imaging remains the clinical mainstay, offering surgeons a natural-color view of tissues [2]. However, certain critical features—such as neoplastic regions or vascular pathways—may be difficult or impossible to distinguish under white light alone [3]. Fluorescence imaging further expands white-light imaging by providing molecular or functional information invisible to the naked eye [4]. Each fluorescence modality supplies additional, complementary insights—such as tumor margins, vascular structures, or lymphatic pathways—to the standard anatomical view [5]. Moreover, multiple targeted fluorophores are currently under clinical investigation, aiming to deliver even more accurate and specific visualization of tissue pathology, function, and anatomy [6].

Although intraoperative fluorescence imaging holds considerable promise, its full clinical value remains underutilized due to persistent challenges in workflow integration [7]. White-light reflectance and fluorescence imaging inherently require distinct illumination and detection spectral bands [8]. Current systems address this need by physically exchanging optical filters when switching between modes. Consequently, only one imaging mode is available at a time, forcing surgeons to view the anatomical perspective and each fluorescence modality sequentially, with a switching delay of several seconds [9]. Yet, each mode provides essential insights—white light for overall anatomical context, fluorescence for delineating abnormal tissue—so neither alone suffices for precise interventions [10]. In neurosurgery, for instance, preserving healthy tissue is as crucial as excising malignant regions, underscoring the need for continuous access to both anatomical and fluorescence views. This sequential workflow demands that surgeons mentally overlay memorized fluorescence data (e.g., PpIX – highlighted cancerous tissue) onto the live white-light view to guide procedures such as tumor resection [11]. In practice, this mental overlay can introduce spatial inaccuracies—owing to imperfect registration of transient imaging data—as well as increased susceptibility to error when intraoperative tissue deformation or resection progress alters visual references [12]. To mitigate these risks, surgeons must switch frequently between imaging modes, especially as tissue appearance changes during surgery [13]. Such a fragmented imaging process not only disrupts surgical flow but also creates cognitive burdens that ultimately impede broader clinical adoption of intraoperative fluorescence imaging [14].

To overcome these limitations and fully harness the potential of advanced optical imaging, an ideal solution would simultaneously capture both reflectance and fluorescence modes in real-time, thereby unifying the advantages of each and preserving the continuous, high-fidelity feedback essential for modern surgery [15]. By merging anatomical and fluorescence data into a single view, surgeons would be spared the need for frequent mode switching or mental image fusion, leading to reduced cognitive load, enhanced spatial accuracy, and better overall procedural focus. This is especially critical in complex surgeries where real-time delineation of tissues and physiological processes can directly affect patient outcomes [16].

Multispectral technologies can capture multiple spectral bands in real-time, yet they remain restricted to one mode—reflectance or fluorescence—at a time [17]. This limitation arises partly because PpIX and fluorescein fluorescence are emitted in visible spectral regions (red and green, respectively), overlapping with standard white-light signals [18]. Time-multiplexed methods [19,20] (e.g., filter-wheel approaches) rapidly alternate between imaging modes, often causing motion artifacts and suffering from slow acquisition rates that compromise real-time feedback [21]. Additionally, such approaches require rapid changes in illumination, which can be visually disruptive in open surgeries and cause stroboscopic effects [22] These constraints prevent current systems from providing continuous multispectral imaging at the spatial and temporal resolutions required in the operating room, revealing a critical need for new approaches that address these technical hurdles.

We introduce a multispectral imaging system tailored to clinical intraoperative workflows. This platform captures, in real-time, white-light reflectance and all three clinically relevant fluorescence modes (PpIX, ICG, and fluorescein) simultaneously [23]. By delivering perfectly spatially and temporally co-registered information in a single, unified view, it removes the need for surgeons to mentally correlate separate images, thus reducing cognitive overhead and minimizing workflow disruptions [24]. Built from readily available components, the system is designed for ease of integration, paving the way for broader and more effective adoption of intraoperative fluorescence imaging.

In the sections that follow, we detail the system's technical setup and demonstrate its ability to image multiple fluorophores alongside standard color reflectance at minimal latency. We then present phantom-based evaluations showcasing accurate fluorophore separation and robust image quality suitable for intraoperative use. Finally, we compare our platform with prior art, illustrating how continuous reflectance–fluorescence imaging can expand clinical capabilities, reduce mental overhead, and preserve an uninterrupted surgical workflow.

## 2. Materials and Methods

### 2.1. Multispectral Imaging System

Our multispectral imaging system was engineered to combine the advantages of a conventional camera—such as simplicity, compactness, absence of moving parts, reliability, low cost, high light sensitivity, and fast operation with enhanced functional imaging capabilities [25]. The system integrates two dedicated sensors, optimized illumination, and optical components to achieve real-time, simultaneous imaging of reflectance and multiple fluorescence signals.

The multispectral imaging system is designed to capture both reflectance and fluorescence images simultaneously, providing comprehensive visual information without disrupting the surgical workflow. The core concept involves splitting the light collected by the imaging optics into two separate optical paths, each leading to a dedicated sensor optimized for specific spectral content. By strategically shaping the illumination spectrum and employing custom optical filters, one sensor captures reflected light in the visible spectrum, while the other sensor detects fluorescence emissions from multiple fluorophores. The fluorescence sensor receives a mixture of emission signals, which are then spectrally resolved using the camera's Bayer filter pattern and an unmixing algorithm. This approach leverages the method of using a color camera to detect and resolve multiple spectral signals [26].

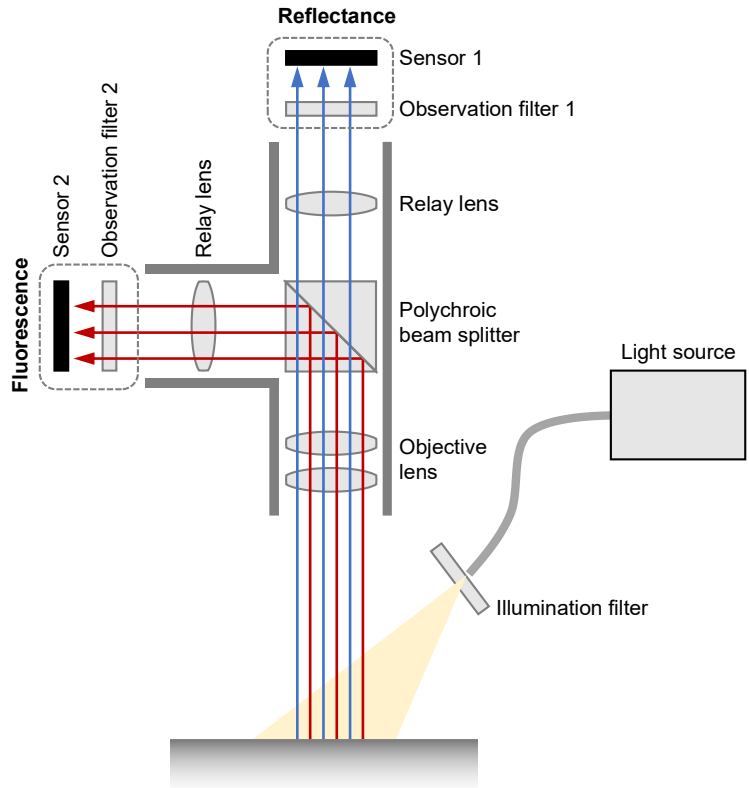

Figure 1: Schematic diagram of the multispectral imaging system, illustrating the set-up of the light source, filters, beam splitter, sensors, and light pathways.

The system comprises a xenon arc lamp, commonly used in surgical microscopes, as its light source. This lamp provides a continuous spectrum covering visible and near-infrared (NIR) wavelengths, delivering the power and spectral characteristics needed to excite multiple fluorophores and provide adequate white light for reflectance imaging. Custom optical filters are employed, including excitation and emission filters tailored to the absorption and emission characteristics of the selected fluorophores.

The illumination filter selectively transmits wavelengths required for fluorophore excitation while ensuring sufficient white light, while the reflectance observation filter matches the illumination filter's transmission bands to optimize reflectance imaging. The fluorescence observation filter features narrow transmission bands aligned with the emission peaks of the fluorophores, enabling simultaneous detection without crosstalk. A custom-designed polychroic beam splitter efficiently separates incoming light based on wavelength, directing reflected light to the reflectance sensor and fluorescence emissions to the fluorescence sensor without significant light loss. This design minimizes optical losses and maximizes signal strength for both imaging modalities. The imaging system incorporates two high-resolution color cameras (Basler), one dedicated to reflectance imaging and the other to fluorescence imaging. The reflectance sensor captures images in the visible spectrum with high spatial fidelity, while the fluorescence sensor, identical in design but without the visible light filter, ensures unrestricted NIR sensitivity. This configuration allows the fluorescence sensor to capture emission signals from multiple fluorophores simultaneously across a broader spectral range.

**Key Innovation:** The unique capability of our system lies in the use of a standard color camera with a Bayer filter [27], combined with custom optical filters and real-time linear unmixing algorithms [28]. This allows for the simultaneous detection and separation of multiple fluorescence signals without the need for complex or expensive equipment. By leveraging this innovative approach, the system provides real-time, integrated visualization essential for surgical precision.

The system employs an illumination filter positioned before the light source to selectively transmit wavelengths required for fluorophore excitation, such as ~400 nm for protoporphyrin IX (PpIX), ~470 nm for fluorescein, and ~785 nm for indocyanine green (ICG). This setup ensures adequate power for excitation, critical for effective fluorescence imaging. The filter also provides sufficient white light tailored to the sensitivity of the sensors, ensuring high-quality reflectance imaging. The reflectance observation filter complements the illumination filter by matching its transmission bands, optimizing reflectance imaging to enhance the signal-to-noise ratio and improve color fidelity. The fluorescence observation filter is designed with narrow transmission bands (40 nm at ~470 nm for fluorescein, 30 nm at ~630 nm for PpIX, and 75 nm at ~830 nm for ICG), enabling simultaneous detection without crosstalk. Additionally, a custom-designed polychroic beam splitter efficiently separates reflectance and fluorescence wavelengths, minimizing optical losses and maximizing signal strength for both imaging modalities.

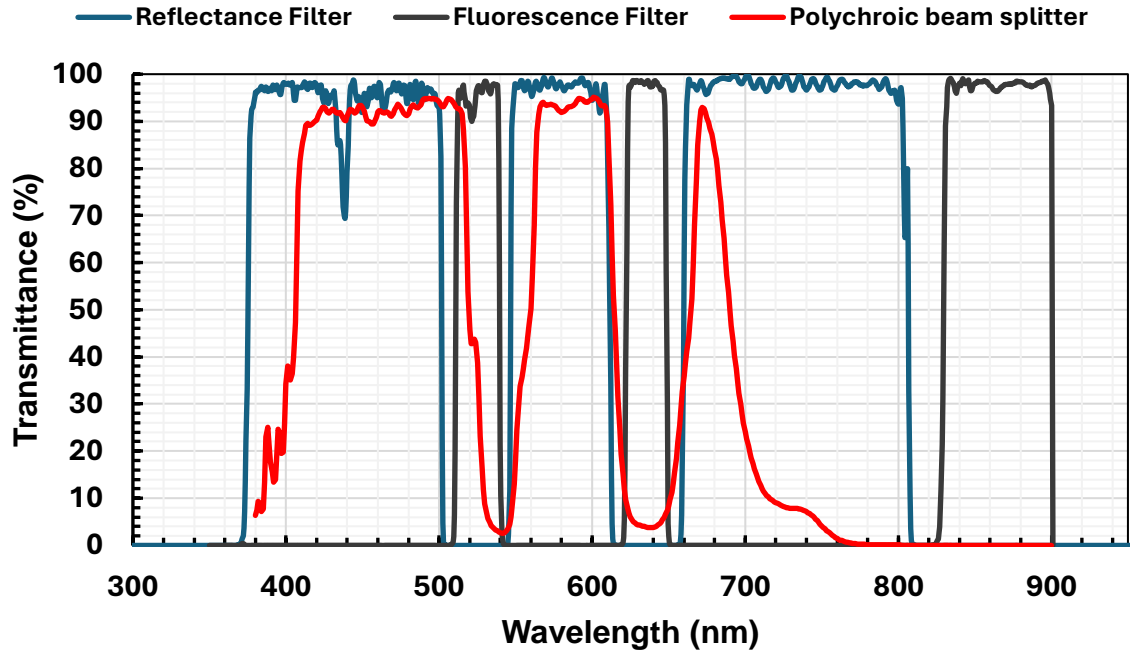

Figure 2: Transmission spectra of the optical components. The fluorescence filter transmits three narrow bands aligned with the peaks of the targeted fluorophores. The reflectance filter spans the visible range, excluding only two narrow bands corresponding to the fluorescence filter's visible fluorescence bands. Custom design of both filters prevents crosstalk. The polychroic beam splitter's transmittance ensures efficient light splitting.

The system includes a reflectance sensor, a color camera (CMOS Image Sensor), that captures reflectance images in the visible spectrum with high spatial fidelity. The fluorescence sensor is identical to the reflectance sensor but lacks the visible light filter, ensuring unrestricted sensitivity in the near-infrared (NIR) range. Both sensors are synchronized using hardware-triggered signals to enable simultaneous image acquisition, ensuring precise temporal alignment between the reflectance and fluorescence images.

### 2.2. Software and Data Processing

Images from both sensors are captured at the same time using hardware triggers, ensuring that the reflectance and fluorescence images are perfectly synchronized. Initially, the two sensors are manually aligned so that their fields of view overlap. A one-time software adjustment is then applied, which automatically keeps the images aligned during use, even if the surgeon changes the magnification or distance. This adjustment uses mathematical transformations, such as shifting, resizing, and rotating the images, to correct any minor misalignments.

To ensure precise calibration, we corrected vignetting—the natural fall-off in intensity toward the edges of the images—caused by illumination characteristics and optical properties. By applying correction factors calculated from reference images taken at standard working distances, we eliminated edge darkening during calibration, resulting in accurate and consistent measurements.

Since the reflectance camera's lighting setup was modified by adding a filter, a recalibration process is necessary to ensure accurate color reproduction. Using a standard color checker chart, the reflectance camera is adjusted by computing a color correction matrix that maps the sensor's raw RGB values to standardized color values.

$$\begin{bmatrix} A_R \\ A_G \\ A_B \end{bmatrix} = \begin{bmatrix} X_{R_1}^{Refl} & X_{R_2}^{Refl} & X_{R_3}^{Refl} \\ X_{G_1}^{Refl} & X_{G_2}^{Refl} & X_{G_3}^{Refl} \\ X_{B_1}^{Refl} & X_{B_2}^{Refl} & X_{B_3}^{Refl} \end{bmatrix} \cdot \begin{bmatrix} S_R^{Refl} \\ S_G^{Refl} \\ S_B^{Refl} \end{bmatrix}$$

$$X^{Refl} = A \cdot (S^{Refl})^{-1}$$

$X^{Refl}$ = Color Correction Matrix for Reflectance
$S^{Refl}$ = Detected signal of the sensor for Reflectance
A = Known value from color checker supplier (ColorGauge Nano Target, Matt, Image Science Associates)

This calibration step is crucial for ensuring the system delivers reliable and precise color imaging, which is essential for detailed intraoperative guidance. To evaluate the color fidelity and accuracy of the system, we use the Delta E (ΔE) metric [29], which quantifies the perceptual difference between the calibrated images and reference values.

The system employs linear unmixing to precisely extract individual fluorescence signals from the captured fluorescence image. This method allows for the accurate identification and differentiation of multiple fluorophores within the field of view, even when their emission spectra overlap. The system is calibrated to detect and quantify three different fluorophores.

Calibration samples with known concentrations of each fluorophore are used to calculate the unmixing matrix. The calibration matrix $X^{fluo}$ is computed using the detected signals from the calibration samples.

$$\begin{bmatrix} C_{f1} \\ C_{f2} \\ C_{f3} \end{bmatrix} = \begin{bmatrix} X_{R_1}^{fluo} & X_{R_2}^{fluo} & X_{R_3}^{fluo} \\ X_{G_1}^{fluo} & X_{G_2}^{fluo} & X_{G_3}^{fluo} \\ X_{B_1}^{fluo} & X_{B_2}^{fluo} & X_{B_3}^{fluo} \end{bmatrix} \cdot \begin{bmatrix} S_R^{fluo} \\ S_G^{fluo} \\ S_B^{fluo} \end{bmatrix}$$

$$X^{fluo} = C_f \cdot (S^{fluo})^{-1}$$

where $S^{fluo}$ is the detected fluorescence signal of the sensor, and $C_f$ represents the estimated concentrations of each fluorophore.

By determining the calibration matrix through careful analysis of known concentration samples, we can apply this matrix to effectively separate the fluorescence signals into distinct components. This approach enhances the reliability of fluorescence imaging, ensuring the surgeon has access to detailed and accurate data during procedures. The use of a color camera with a Bayer filter pattern to detect and resolve multiple spectral signals [30].

To present the fluorescence information in a way that is intuitive for the surgeon, we employ a technique called pseudocoloring [31]. This process assigns distinct, easily recognizable colors to the fluorescence signals, ensuring they are clearly differentiated from the natural reflectance image and minimizing potential confusion. The choice of colors is flexible and can be adjusted based on the specific application or user preference. In this study, the following colors were assigned as an example:

R630 (Phosphor Tech) (simulating PpIX induced fluorescence [32]): Blue

Fluorescein sodium [33]: Green

Indocyanine green [34]: Red

Once the fluorescence data is pseudocolored, the images—one representing reflectance and the other fluorescence—are combined into a single composite image that retains all critical information. This integration occurs in real-time, allowing the surgeon to visualize both anatomical structures and fluorescence signals simultaneously.

The combination of the two images can be achieved through either linear or non-linear processes. In the case of linear blending, the images are overlaid by summing the pixel values, ensuring that both reflectance and fluorescence information are clearly visible. This straightforward method provides a comprehensive, enhanced view of the surgical field, conveying multiple pieces of information beyond conventional color imaging or fluorescence-only images.

### 2.3. Phantom Preparation

To simulate clinical scenarios and validate the system, we prepared phantoms [35] using fluorophores that mimic the optical properties of those used in surgical applications. The simulation of Protoporphyrin IX (PpIX) fluorescence, resulting from 5-ALA administration [36], was achieved using the R630 dye (Phosphor Tech). A total of 4 mg of the dye was dissolved in 2 mL of distilled water. The dye exhibited an excitation peak at approximately 400 nm and an emission peak at 630 nm.

Fluorescein sodium was prepared by dissolving 1 mg of the compound in 8 mL of distilled water. This fluorophore exhibited an excitation peak at approximately 470 nm and an emission peak at 520 nm.

Indocyanine Green (ICG) was prepared by dissolving 1 mg of the dye and 2 g of albumin in 8 mL of distilled water to mimic plasma binding. ICG exhibited an excitation peak at approximately 785 nm and an emission peak at 830 nm.

Each fluorophore solution was placed in separate transparent vials and arranged in various configurations to test the system's ability to detect and differentiate multiple fluorophores.

### 2.4. Imaging System Settings

The surgical microscope used allows the user to select the working distance and magnification according to the surgical requirements. For this study, we selected specific settings to standardize the experiments, but these settings do not affect the presented technology or principle. Key settings and parameters used during the experiments are summarized below:

Field of View: 9.7 mm (W) × 7.5 mm (H)

Working Distance: 350 mm

Magnification: 1.4×

Sensor Resolution: 1920 × 1200 pixels; Pixel size: 5.86 μm²

Processing Software: MATLAB 2022a

## 3. Results

### 3.1. Simultaneous Fluorescence Imaging

To evaluate the system's capability for simultaneous multispectral fluorescence imaging, we conducted the measurements using phantoms prepared as described in Section 2.3. Four transparent vials were filled with fluorophore solutions: three vials each containing a single fluorophore—R630, fluorescein sodium, or indocyanine green (ICG)—and a fourth vial containing a 50:50 mixture of R630 and fluorescein sodium.

The fluorescence sensor captured the combined emission signals from all fluorophores simultaneously. Utilizing the multispectral capabilities of the system—achieved through the combination of hardware components (color camera with Bayer filter array and triple-bandpass filters) and the linear unmixing algorithm—we were able to resolve overlapping fluorescence emissions into separate spectral components in real-time.

Figure 3(A)-(C) display the raw fluorescence image captured by the fluorescence sensor, where the emission signals from different fluorophores overlap. After applying the linear unmixing algorithm, distinct fluorescence images for each fluorophore were obtained, as shown in Figure 3(D)–(F):

(D): Unmixed fluorescence image of fluorescein sodium.

(E): Unmixed fluorescence image of ICG.

(F): Unmixed fluorescence image of R630.

These images demonstrate the system's ability to accurately separate the fluorescence signals of fluorescein sodium, ICG and R630 without significant crosstalk or background noise.

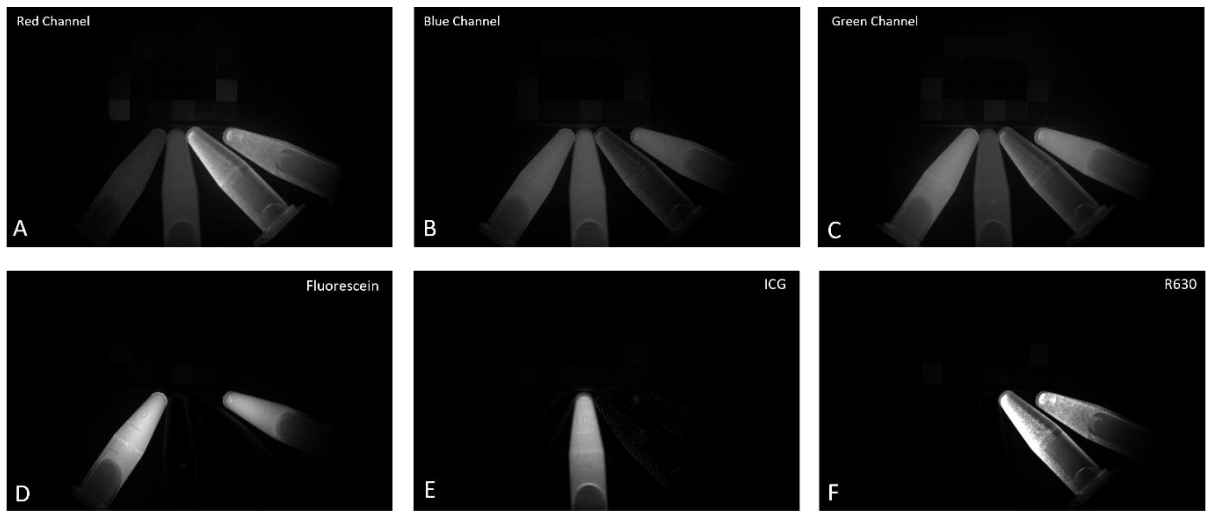

Figure 3: Raw fluorescence image displaying combined signals of the three fluorophores and the unmixed fluorescence images for (D) fluorescein, (E)ICG, and (F)R630.

In the pseudocolored composite fluorescence image (Figure 4), each fluorophore is assigned a distinct color—blue for R630, green for fluorescein sodium, and red for ICG. The fourth vial containing the mixture of R630 and fluorescein sodium appears cyan due to the additive mixing of blue and green colors. This result confirms the successful detection and differentiation of multiple fluorophores within the same field of view, highlighting the system's capability to provide comprehensive fluorescence information in real-time.

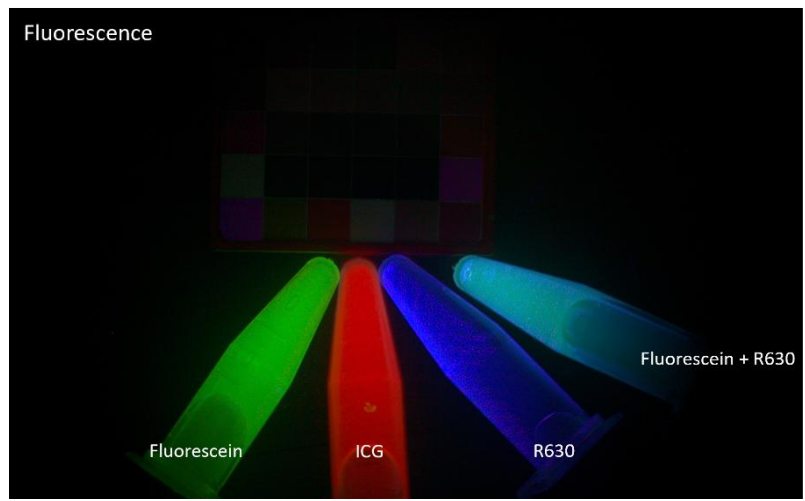

Figure 4: Pseudocolored fluorescence image illustrating the spatial distribution and overlap of fluorophores.

### 3.2. Reflectance Imaging

To assess the color fidelity of the reflectance imaging, we performed a quantitative evaluation using a standard color checker chart. The calibrated reflectance image obtained from our system was compared with a reference image captured by a conventional color camera under identical lighting conditions.

Figure 5 shows the comparison between (A) the calibrated reflectance image from our system and (B) the reference image from the commercial camera of White light mode. Visually, there is no noticeable color difference for most of the color patches. For patches where minor color differences exist, the discrepancies are minimal and comparable to the typical variations observed between different color cameras.

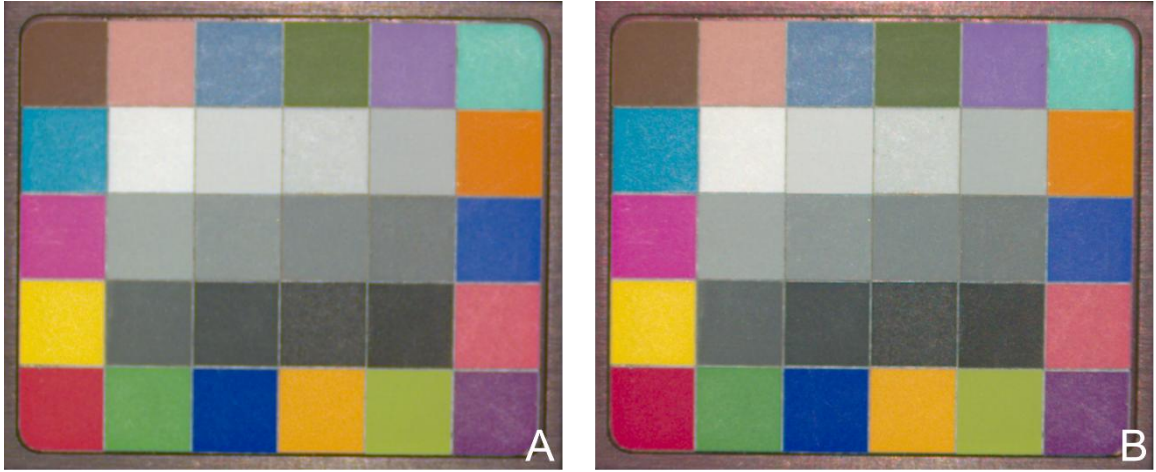

Figure 5: Comparison between (A) calibrated reflectance multispectral image and (B) reference image from microscope standard color imaging mode.

We quantified the color differences using the Delta E (ΔE) metric, a widely accepted measure of perceptual color difference. According to established guidelines, a ΔE value less than 1 is considered imperceptible to the human eye, while values between 1 and 2 are barely perceptible [37].

Figure 6 presents the ΔE values for each color patch on the chart. The majority of the color patches exhibited ΔE values below 2, indicating high color accuracy. Specifically:

ΔE < 1: 65% of the patches, imperceptible difference.

1 ≤ ΔE < 2: 25% of the patches, barely perceptible difference [37].

ΔE ≥ 2: 10% of the patches, noticeable but acceptable difference.

These results demonstrate that despite the exclusion of two narrow visible spectral bands from the color acquisition process (due to the filters used for fluorescence emission), the system maintains excellent color fidelity in the reflectance images. The acceptable level of color inaccuracies confirms that the modifications made to enable multispectral fluorescence imaging do not adversely affect the quality of anatomical visualization.

By ensuring high color accuracy, the system provides reliable reflectance images essential for detailed anatomical assessment during surgical procedures.

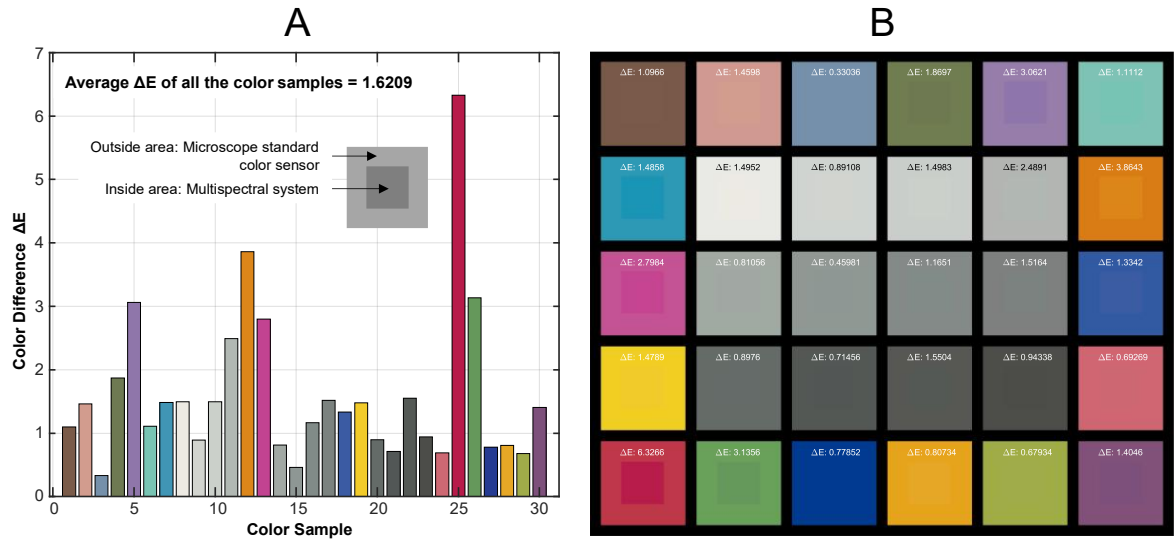

Figure 6: (A) Color differences ΔE for the color checker chart patches measured by our multispectral imaging system compared to the microscope standard color imaging mode, (B) Visualization of ΔE values for the color checker chart patches with corresponding color squares.

### 3.3. Composite Imaging

The integration of the calibrated reflectance image with the unmixed fluorescence images resulted in a composite image that provides comprehensive visual information in a single, unified view. This composite image allows for the simultaneous visualization of anatomical structures and fluorescence data, enhancing the surgeon's ability to interpret the scene intuitively and make informed decisions in real time.

Figure 7 illustrates the composite image displaying the color checker chart alongside the three fluorescence signals visualized in their assigned pseudocolors—blue for R630, green for fluorescein sodium, and red for ICG. The composite image clearly shows the accurate alignment of the reflectance and fluorescence images, as well as the distinct localization of each fluorophore. The areas where fluorophores overlap are represented by additive color mixing, such as cyan indicating the presence of both R630 and fluorescein sodium in the same region.

This result demonstrates the system's capability to seamlessly combine anatomical and fluorescence imaging, providing a more informative and intuitive visualization compared to viewing each modality separately. By delivering all relevant visual information in a single image, the system reduces cognitive load on the surgeon and enhances spatial awareness during surgical procedures.

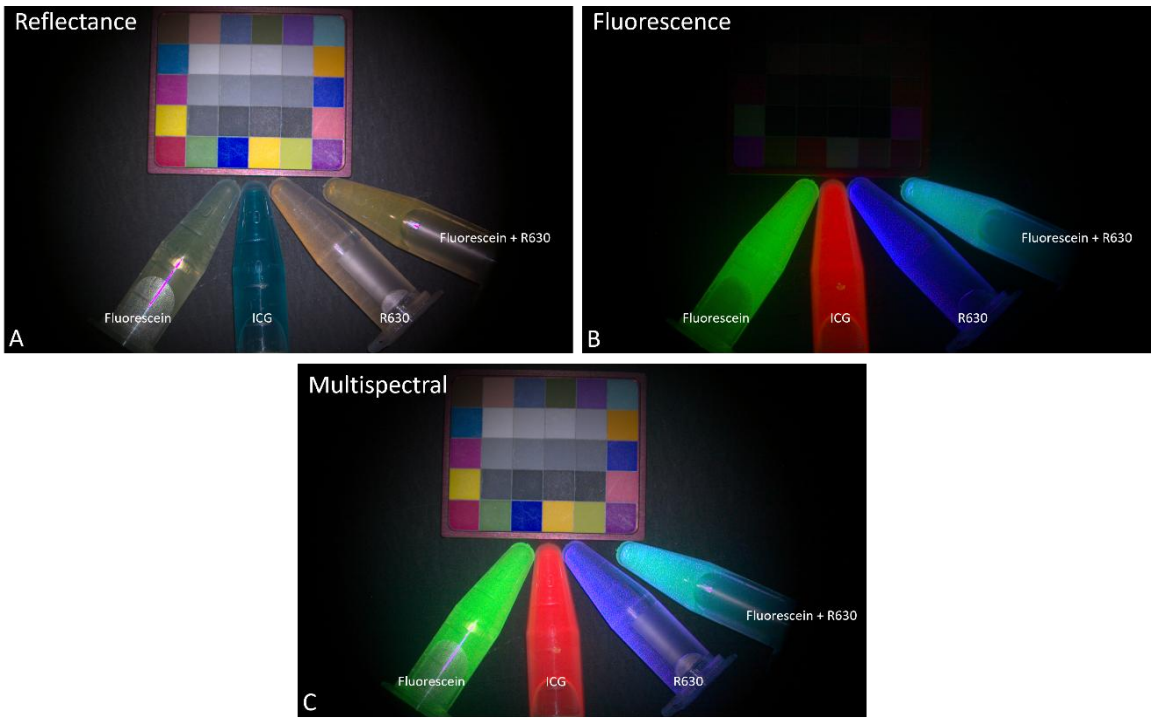

Figure 7: (A) Reflectance image captured by the designed reflectance sensor, showing the anatomical information. (B) Fluorescence image captured by the fluorescence sensor, visualized in pseudocolors: blue for R630, green for fluorescein sodium, and red for ICG. (C) Composite image combining the calibrated reflectance and unmixed fluorescence images, demonstrating accurate alignment and additive color mixing for overlapping fluorophores (e.g., cyan for fluorescein sodium and R630).

### 3.4. Real-Time Operation

The system demonstrated real-time performance suitable for intraoperative use, achieving a frame rate of 30 frames per second (FPS) and processing latency of less than 50 milliseconds, ensuring immediate visual feedback for surgeons. Many hyperspectral imaging systems, while offering detailed spectral data, operate at higher latencies due to the computational demands of processing large datasets, making them less practical for dynamic surgical environments. Similarly, time-multiplexed systems, which sequentially switch between imaging modes, often introduce motion artifacts and delays that disrupt workflow. In contrast, our system provides simultaneous reflectance and fluorescence imaging, avoiding these limitations and ensuring spatial and temporal coherence. With its minimal latency and high frame rate, the system offers an optimal balance of speed, accuracy, and real-time functionality critical for surgical applications.

Throughout continuous imaging tasks, the system operated smoothly without any drop in processing speed, frame rate, image resolution, or fluorescence unmixing accuracy. We conducted extended tests simulating surgical procedures to verify stability and reliability, confirming that the system consistently maintained these key performance indicators over time, even under conditions mimicking dynamic surgical environments. The real-time processing of both reflectance and multiple fluorescence signals validates the system's practicality and ergonomic benefits for surgical applications. In dynamic surgical environments where timely decision-making is critical, the ability to provide instantaneous, comprehensive visual information supports better outcomes and enhances overall efficiency.

By eliminating the need for mode switching, our system has the potential to reduce intraoperative delays. While specific time savings are dependent on the surgical procedure, the continuous access to integrated imaging could enhance surgical efficiency and potentially shorten operating times. Further studies are warranted to quantify these benefits.

### 4. Discussion

In this section, we interpret our findings in the context of existing technologies and clinical needs. We discuss how our multispectral imaging system addresses the limitations of current intraoperative imaging solutions and explore its potential impact on surgical practice and patient outcomes.

Our results confirm that the developed multispectral imaging system effectively addresses the limitations of existing intraoperative imaging technologies by overcoming the fragmentation of information. By simultaneously capturing and integrating multiple fluorescence signals with white light imaging, the system provides a unified, comprehensive view of both anatomical and fluorescence information in real-time. This integrated visualization eliminates delays, reduces cognitive burden, and streamlines the surgical workflow.

Technically, the system avoids motion artifacts and visual discomfort associated with time-multiplexed imaging methods that rapidly alternate between imaging modes, often causing stroboscopic effects [22]. These methods typically operate at low frame rates for each mode due to the repeated switching of LEDs on and off, which is inadequate for dynamic surgical environments requiring real-time feedback. Our system achieves a frame rate well above the real-time requirements (i.e. 30 FPS) for simultaneous reflectance and fluorescence imaging, influenced by the intensity of the illumination. This frame rate can exceed levels typically considered real-time, ensuring smooth visualization without temporal gaps.

The integration of multiple fluorescence signals with white light imaging enhances visual clarity and spatial accuracy, supporting more precise surgical interventions. The system allows surgeons to visualize anatomical structures and fluorescence information simultaneously, providing a comprehensive view that was not previously possible with conventional systems. This enhanced visualization facilitates better intraoperative decision-making, potentially leading to improved patient outcomes.

By providing continuous, simultaneous, and integrated visualization of both anatomical and fluorescence information, the system streamlines the surgical workflow. Surgeons receive comprehensive real-time data without delays or the need to mentally integrate fragmented information from separate images. This elimination of fragmentation reduces cognitive load and the potential for errors, allowing surgeons to maintain focus on the operative field and enhance precision.

Improved surgical accuracy can potentially reduce operating times and decrease complication rates, contributing to better patient outcomes and faster recoveries. These benefits may translate into cost savings for healthcare facilities, making the technology economically advantageous.

Additionally, the system's capability to capture multiple fluorescence signals opens avenues for advanced multi-fluorophore analysis. For instance, the collective evaluation of fluorescence signals and tissue color can enhance image segmentation and tissue characterization. This could lead to more insightful assessments of tumor vascularization or tissue perfusion, allowing surgeons to optimize interventions with greater accuracy and safety—such as ensuring that the excision of tumorous tissue does not adversely impact neighboring healthy brain tissue. The flexible design of the system also ensures compatibility with new fluorophores that may be developed in the future, extending its range of applications in surgical imaging.

Although the presented configuration of the multispectral system is a clear advancement over existing systems and meets current clinical practice needs, there is still room for further studies and technological improvements. As new fluorescence agents will be developed, there may be a need to adapt the fluorescence channels to different spectral bands. Enhancing the system's adaptability to accommodate a wider range of fluorophores would increase its flexibility in various applications.

Moreover, this study utilized phantoms to simulate clinical conditions. Clinical validation in real surgical environments is necessary to assess performance under conditions such as tissue autofluorescence, varying optical properties, and the presence of blood or other fluids. Comprehensive clinical studies quantifying the system's impact on patient outcomes are needed to substantiate its benefits fully.

Future work includes conducting in vivo clinical trials to validate the system's effectiveness and quantify its impact on surgical outcomes. We plan to explore the adaptability of the system to new fluorophores, enhancing its flexibility for various surgical applications. We compare our system with multiplexing technology because both methods aim to enhance intraoperative imaging by capturing multiple spectral signals. However, multiplexing introduces mechanical complexity and latency, whereas our system ensures real-time, simultaneous visualization with minimal disruption to surgical workflows. This distinction highlights the advantages of our approach in clinical settings.

**Table 1: SWOT analysis of Multispectral Imaging System vs Multiplexing Technology**

| Multispectral Imaging System | Multiplexing Technology |
| --- | --- |

| | | |
|---|---|---|
| Strength | Provides simultaneous, real-time visualization of reflectance and fluorescence without mode switching. | Enables real-time imaging by alternating modalities at high speed. |
| | No moving parts, leading to enhanced reliability and lower maintenance needs. | Offers superior spectral resolution compared to simpler systems. |
| | High spatial and temporal resolution (30 FPS with <50 ms latency). | |
| | Integrated design reduces cognitive load and streamlines surgical workflows. | |
| | Adaptable for new fluorophores and applications. | |
| Weaknesses | Requires calibration for specific fluorophores, the system can be easily adapted to new agents as they become clinically relevant. | Mechanically complex due to reliance on moving parts (e.g., filter wheels, tunable filters). |
| | | Prone to motion artifacts, mechanical failures, and stroboscopic effects, which can cause visual discomfort, especially during long surgeries. |
| | | Synchronization demands can increase computational load and potential delays. |
| Opportunities | Expanding clinical applications through the integration of new fluorophores. | Hybrid designs combining Multispectral Imaging System with more robust components could enhance reliability. |
| | Potential to improve surgical precision and reduce operating times, leading to widespread adoption. | Potential for niche applications where very high spectral resolution is required. |
| | Future advancements could lower medical costs by reducing surgical duration and improving efficiency. | |
| Threats | Real-time processing may be challenged by increasing computational demands, especially in low-light conditions requiring advanced algorithms. | Increasing preference for simpler, more reliable systems like Multispectral Imaging System. |
| | | Limited ability to capture high-resolution fluorescence and reflectance simultaneously due to system constraints. |
| | | Maintenance challenges because of moving parts could lead to reduced trust and adoption in critical surgical environments. |

## 5. Conclusion

In this study, we introduced a multispectral imaging system that addresses the critical limitations of current intraoperative technologies by enabling real-time, integrated visualization of anatomical and fluorescence information. By unifying white light reflectance with multiple fluorescence signals, the system eliminates mode-switching delays, reduces cognitive load, and streamlines surgical workflows. These advancements not only enhance surgical precision but also hold promise for reducing operating times and improving patient outcomes.

The system's adaptability to future fluorophores and cost-effective design ensures its long-term viability and broad applicability across various surgical fields. Immediate steps include clinical validation through in vivo trials to assess its performance under realistic surgical conditions, such as the presence of tissue autofluorescence, varying optical properties, and interference from blood or fluids.

To further improve this technology, several advancements are envisioned:
Expanding the system's spectral range and compatibility to accommodate emerging fluorophores and new clinical applications.

Investigating compact and portable designs to make the system accessible for broader surgical environments, including resource-limited settings.

Exploring advanced multi-fluorophore analysis techniques for detailed tissue characterization and segmentation, optimizing tumor detection and vascular assessments.

By aligning cutting-edge technology with clinical needs and addressing these future directions, this innovation sets the stage for transformative improvements in surgical practice, paving the way for precision and efficiency in every operation.

## 6. Funding

No funding was received.

## 7. Acknowledgments

The authors thank Leica Microsystems (Heerbrugg, Switzerland) for access to the ARveo 8 surgical microscope and related laboratory resources, and for institutional support and permission to publish this work. The authors also appreciate the technical discussions and practical assistance provided by their Leica colleagues during system development and testing.

## 8. Disclosures

The authors declare no conflicts of interest relevant to this study.

## 9. Data availability statement

The Raw Data presented herein are available on figshare (https://doi.org/10.6084/m9.figshare.28732040)

## 10. References

1. N. J. Harlaar, G. M. van Dam, and V. Ntziachristos, "Intraoperative Optical Imaging," Springer, New York, NY (2014) [doi:10.1007/978-1-4614-7657-3_16].
   https://doi.org/10.1007/978-1-4614-7657-3_16

2. D. Athanasopoulos, A. Heimann, M. Nakamura et al. "Real-Time Overlapping of Indocyanine Green-Video Angiography With White Light Imaging for Vascular Neurosurgery: Technique, Implementation, and Clinical Experience," Oper. Neurosurg. 19 (4), 453–460 (2020) [doi:10.1093/ons/opaa050].
   https://doi.org/10.1093/ons/opaa050

3. E. de Boer, N. J. Harlaar, A. Taruttis et al. "Optical innovations in surgery," Br. J. Surg. 102 (2) (2015) [doi:10.1002/bjs.9713].
   https://doi.org/10.1002/BJS.9713

4. A. A. Cohen-Gadol, D. W. Roberts, and C. G. Hadjipanayis, "Evolving applications of fluorescence technology in neurosurgery," Neurosurg. Focus 36(2) (2014) [doi:10.3171/2013.12.FOCUS13582].
   https://doi.org/10.3171/2013.12.FOCUS13582

5. R. K. Orosco, R. Y. Tsien, and Q. T. Nguyen, "Fluorescence Imaging in Surgery," IEEE Rev. Biomed. Eng. 6(6), 178–187 (2013) [doi:10.1109/RBME.2013.2240294].
   https://doi.org/10.1109/RBME.2013.2240294

6. P. L. Choyke and H. Kobayashi, "Medical Uses of Fluorescence Imaging: Bringing Disease to Light," IEEE J. Sel. Top. Quantum Electron. 18(3), 1140–1146 (2012) [doi:10.1109/JSTQE.2011.2164900].
   https://doi.org/10.1109/JSTQE.2011.2164900

7. O. Bin-Alamer, H. Abou-Al-Shaar, Z. C. Gersey et al. "Intraoperative Imaging and Optical Visualization Techniques for Brain Tumor Resection: A Narrative Review," Cancers 15 (19), 4890 (2023) [doi:10.3390/cancers15194890].
   https://doi.org/10.3390/cancers15194890

8. D. J. Waterhouse, A. S. Luthman, and S. E. Bohndiek, "Spectral band optimization for multispectral fluorescence imaging," 10057, 1005709 (2017) [doi:10.1117/12.2253069].
   https://doi.org/10.1117/12.2253069

9. P. A. Valdés, F. Leblond, V. L. Jacobs et al. "Quantitative, spectrally-resolved intraoperative fluorescence imaging," Sci. Rep. 2 (1), 798 (2012) [doi:10.1038/srep00798].
   https://doi.org/10.1038/SREP00798

10. P. A. Valdés, J. P. Angelo, and S. Gioux, "Real-time quantitative fluorescence imaging using a single snapshot optical properties technique for neurosurgical guidance," 9305, 72–77 (2015) [doi:10.1117/12.2085051].
    https://doi.org/10.1117/12.2085051

11. T. Nagaya, Y. A. Nakamura, P. L. Choyke et al. "Fluorescence-Guided Surgery," Front. Oncol. 7, 314 (2017) [doi:10.3389/fonc.2017.00314].
    https://doi.org/10.3389/FONC.2017.00314

12. M. Rogers, R. I. Cook, R. H. Bower et al. "Barriers to implementing wrong site surgery guidelines: a cognitive work analysis," IEEE Trans. Syst. Man Cybern. A 34 (6), 757–763 (2004) [doi:10.1109/TSMCA.2004.836805].
    https://doi.org/10.1109/TSMCA.2004.836805

13. D. Kuhnt, M. H. A. Bauer, and C. Nimsky, "Multimodality Navigation in Neurosurgery," Springer, New York, NY, pp. 497–506 (2014) [doi:10.1007/978-1-4614-7657-3_36].
https://doi.org/10.1007/978-1-4614-7657-3_36

14. T. J. A. Snoeks, P. B. A. A. van Driel, S. Keereweer et al. "Towards a Successful Clinical Implementation of Fluorescence-Guided Surgery," Mol. Imaging Biol. 16 (2), 147–151 (2014) [doi:10.1007/S11307-013-0707-Y].
https://doi.org/10.1007/S11307-013-0707-Y

15. Dimitriadis N, Grychtol B, Maertins L et al. "Simultaneous real-time multicomponent fluorescence and reflectance imaging method for fluorescence-guided surgery." Optics Letters 41(6):1173-1176 (2016). [doi:10.1364/OL.41.001173].
https://opg.optica.org/ol/abstract.cfm?uri=ol-41-6-1173

16. S. B. Mondal, S. Gao, N. Zhu et al. "Real-time Fluorescence Image-Guided Oncologic Surgery," Vol. 124, 171–211 (2014) [doi:10.1016/B978-0-12-411638-2.00005-7].
https://doi.org/10.1016/B978-0-12-411638-2.00005-7

17. Z. T. Harmany, X. Jiang, and R. Willett, "The value of multispectral observations in photon-limited quantitative tissue analysis," IEEE Signal Process. Workshop Stat. Signal Process., 237–240 (2012) [doi:10.1109/SSP.2012.6319670].
https://doi.org/10.1109/SSP.2012.6319670

18. H. Taniguchi, N. Kohira, T. Ohnishi et al. "Improving Convenience and Reliability of 5-ALA-Induced Fluorescent Imaging for Brain Tumor Surgery," Springer, Cham, pp. 209–217 (2015) [doi:10.1007/978-3-319-24574-4_25].
https://doi.org/10.1007/978-3-319-24574-4_25

19. R. M. Levenson, D. T. Lynch, H. Kobayashi et al. "Multiplexing with multispectral imaging: From mice to microscopy," ILAR J. 49 (1), 78–88 (2008) [doi:10.1093/ilar/49.1.78].
https://doi.org/10.1093/ilar.49.1.78

20. C. Andreou, R. Weissleder, and M. F. Kircher, "Multiplexed imaging in oncology," Nat. Biomed. Eng. 6(5), 527–540 (2022) [doi:10.1038/s41551-022-00891-5].
https://doi.org/10.1038/s41551-022-00891-5

21. G. Bub, M. Tecza, M. Helmes et al. "Temporal pixel multiplexing for simultaneous high-speed, high-resolution imaging," Nat. Methods 7 (3), 209–211 (2010) [doi:10.1038/NMETH.1429].
https://doi.org/10.1038/NMETH.1429

22. A. Sankaranarayanan, L. Xu, C. Studer et al. "Video Compressive Sensing for Spatial Multiplexing Cameras Using Motion-Flow Models," SIAM J. Imaging Sci. 8 (2015) [doi:10.1137/140983124].
https://doi.org/10.1137/140983124

23. G. Themelis, A. Sarantopoulos, N. J. Harlaar et al. "Real-time intra-operative fluorescence imaging with targeted fluorophores," (2010) [doi:10.1364/BIOMED.2010.BWD6].
https://doi.org/10.1364/BIOMED.2010.BWD6

24. R.-J. Swijnenburg, L. M. A. Crane, B. Kt et al. "Intraoperative imaging using fluorescence," Ned. Tijdschr. Geneeskd. 156 (11) (2012).
https://pubmed.ncbi.nlm.nih.gov/22414671/

25. R. V. Smith, "The digital camera in clinical practice," Otolaryngol. Clin. North Am. (2002) [doi:10.1016/S0030-6665(02)00066-X].
https://doi.org/10.1016/S0030-6665(02)00066-X

26. G. Themelis, J. S. Yoo, K.-S. Soh et al. "Real-time intraoperative fluorescence imaging system using light-absorption correction," J. Biomed. Opt. 14 (6), 064012 (2009) [doi:10.1117/1.3259362].
https://doi.org/10.1117/1.3259362

27. R. Lukac, K. N. Plataniotis, and A. N. Venetsanopoulos, "Bayer pattern demosaicking using local-correlation approach," in Advances in Imaging Technology (2004).
https://doi.org/10.1007/978-3-540-25944-2_4

28. G. Themelis, J. S. Yoo, and V. Ntziachristos, "Multispectral imaging using multiple-bandpass filters," Opt. Lett. 33, 1023–1025 (2008).
https://doi.org/10.1364/OL.33.001023

29. S. P. Farnand, "Using [Delta]E metrics for measuring color difference in hard copy pictorial images," Proc. SPIE, Color Imaging VIII: Processing, Hardcopy, and Applications 5008, (2003) [doi:10.1117/12.474873].
https://doi.org/10.1117/12.474873

30. G. Themelis, N. J. Harlaar, W. Kelder et al., "Enhancing Surgical Vision by Using Real-Time Imaging of αvβ3-Integrin Targeted Near-Infrared Fluorescent Agent," *Ann. Surg. Oncol.* 18, 3506–3513 (2011) [doi:10.1245/s10434-011-1664-9].
https://doi.org/10.1245/s10434-011-1664-9

31. J. Glatz, P. Symvoulidis, P. B. Garcia-Allende et al. "Robust overlay schemes for the fusion of fluorescence and color channels in biological imaging," J. Biomed. Opt. 19 (4), 040501 (2014) [doi:10.1117/1.jbo.19.4.040501].
   https://doi.org/10.1117/1.jbo.19.4.040501

32. S. A. Goryaynov et al., "The role of 5-ALA in low-grade gliomas and the influence of antiepileptic drugs on intraoperative fluorescence," Front. Oncol. (2019) [doi:10.3389/fonc.2019.00423].
   https://doi.org/10.3389/FONC.2019.00423

33. B. Küçükyürük et al., "Intraoperative fluorescein sodium videoangiography in intracranial aneurysm surgery," World Neurosurg. 147, e444–e452 (2021) [doi:10.1016/j.wneu.2020.12.085].
   https://doi.org /10.1016/j.wneu.2020.12.085

34. E. H. Kim et al., "Application of intraoperative indocyanine green videoangiography to brain tumor surgery," Acta Neurochir. (2011) [doi:10.1007/S00701-011-1046-X].
   https://doi.org/10.1007/S00701-011-1046-X

35. D. Gorpas, M. Koch, M. Anastasopoulou et al. "Benchmarking of fluorescence cameras through the use of a composite phantom," J. Biomed. Opt. 22 (1), 016009 (2017) [doi:10.1117/1.JBO.22.1.016009].
   https://doi.org/10.1117/1.JBO.22.1.016009

36. P. A. Valdés, F. Leblond, A. Kim et al. "ALA-induced PpIX spectroscopy for brain tumor image-guided surgery," Proc. SPIE 7883, Photonic Therapeutics and Diagnostics VII (2011) [doi:10.1117/12.875016].
   https://doi.org/10.1117/12.875016

37. K. G. Ghassemian, P. A. Bain, S. Salari et al. "Perceptibility and acceptability thresholds for colour differences in dentistry," J. Dent. 42 (6), 637–644 (2014) [doi:10.1016/j.jdent.2013.11.017].
   https://doi.org/10.1016/j.jdent.2013.11.017

**Author Biographies**

**Kyriakos Pentarakis** is a PhD candidate at the Hellenic Mediterranean University and an Innovation Scientist at the Advanced Technology Lab of Leica Microsystems. He holds a BSc in Physics and an MSc in Medical Optics and Vision from University of Crete. His research focuses on developing innovative optical imaging applications for microsurgery, with an emphasis on integrating fluorescence and multispectral imaging technologies into surgical workflows.

**Dr. Konstantinos Petridis** is a Professor in the Department of Electronic Engineering at the Hellenic Mediterranean University (HMU) and the HMU Vice Rector of Internationalization & Extroversion since 2023. He obtained his Ph.D. in physics from the University of St. Andrews (UK) in 2002. His research focuses on applications of lasers in materials processing and their applications in third-generation solar cells and gas sensing. He demonstrates expertise and extended laboratory experience in laser technology and physics, as well as the ultrafast laser processing techniques of graphene and graphene-based optoelectronic devices and laser-induced decoration of 2D materials with nanomaterials. He has coordinated 12 Erasmus and Erasmus Plus projects, and he actively participated as a teacher, manager, and sub-coordinator in another 12 Erasmus projects. He has participated as a researcher in 12 national & international research programs. His research has received over 2200 citations, and his h-index is 25.

**Dr. George Kakavelakis** is an Assistant Professor at Hellenic Mediterranean University. He earned his PhD in 2018 from the University of Crete (IKY-Siemens Scholar), with research focused on solution-processed photovoltaics. He was a Research Fellow at the University of Cambridge, working on printed optoelectronic devices. In 2021, he joined EPFL as a Marie Skłodowska-Curie Fellow to develop low-cost energy devices. His current research focuses on the development of novel optoelectronic devices and advanced characterization techniques.

**Dr. George Themelis** received his Ph.D. in Medical Physics from the University of Crete, Greece, in 2002, specializing in spectral imaging methods for early detection and staging of epithelial malignancies. His research career includes positions as a Research Fellow at Harvard Medical School and Massachusetts General Hospital, Head of the Medical Imaging Laboratory at Helmholtz Zentrum München, and Advanced Technology Manager at Leica Microsystems (Switzerland). Dr. Themelis currently leads research and development in optical imaging as the founder of Theta Photonics UG (Germany). His primary contributions are in multispectral optical imaging techniques and intraoperative fluorescence imaging, with innovations translated into clinical and commercial medical devices.